\documentclass [prb,twocolumn,superscriptaddress,preprintnumbers]{revtex4}

\usepackage {graphicx}
\usepackage {amsmath}
\usepackage {bm}
\usepackage {subfigure}

\begin{document}

\title{
Ultrafast, ultrabright, X-ray holography using a uniformly-redundant array
}

\begin{abstract}
 Advances in the development of free-electron lasers offer the
realistic prospect of high-resolution imaging to study the nanoworld
on the time-scale of atomic motions. We identify X-ray Fourier
Transform holography\cite{stroke,mcnulty,eisebitt}, (FTH) as a
promising but, so far, inefficient scheme to do this. We show that a
uniformly redundant array\cite{fenimore:URA} (URA) placed next to the
sample, multiplies the efficiency of X-ray FTH by more than one
thousand (approaching that of a perfect lens) and provides holographic
images with both amplitude- and phase-contrast information.  The
experiments reported here demonstrate this concept by imaging a
nano-fabricated object at a synchrotron source, and a bacterial cell
at a soft X-ray free-electron-laser, where illumination by a single 15
fs pulse was successfully used in producing the holographic image. We
expect with upcoming hard X-ray lasers to achieve considerably higher
spatial resolution and to obtain ultrafast movies of excited states of
matter.
\end{abstract}

\author{Stefano ~Marchesini}
\affiliation{Lawrence Livermore National Laboratory, 7000 East
Ave., Livermore, CA 94550, USA.}
\affiliation{Advanced Light Source, Lawrence Berkeley National
Laboratory, 1 Cyclotron rd. Berkeley, CA 94720, USA}
\email[Correspondance should be addressed to:]{smarchesini@lbl.gov}

\author{S\'ebastien Boutet}
\affiliation{Stanford Synchrotron Radiation Laboratory, 
Stanford Linear Accelerator Center, 2575 Sand Hill Road, Menlo Park, 
California 94025, USA.}
\affiliation{Laboratory of Molecular Biophysics, 
Department of Cell and Molecular Biology, 
Uppsala University, Husargatan 3, Box 596, SE-75124 Uppsala, Sweden.}
\author{Anne E. Sakdinawat}
\affiliation{Center for X-ray Optics, Lawrence Berkeley National Laboratory, Berkeley, California 94720, USA.}

\author{Michael J. Bogan}
\affiliation{Lawrence Livermore National Laboratory, 7000 East
Ave., Livermore, CA 94550, USA.} 
\author{Sa\u sa Bajt}
\affiliation{Lawrence Livermore National Laboratory, 7000 East
Ave., Livermore, CA 94550, USA.}
\author{Anton Barty}  
\affiliation{Lawrence Livermore National Laboratory, 7000 East
Ave., Livermore, CA 94550, USA.}
\author{Henry N. Chapman} 
\affiliation{Lawrence Livermore National Laboratory, 7000 East
Ave., Livermore, CA 94550, USA.}
\affiliation{
Centre for Free-Electron Laser Science 
U. Hamburg, DESY,
Notkestra\ss e 85, Hamburg, Germany.
}
\author{Matthias Frank}
\affiliation{Lawrence Livermore National Laboratory, 7000 East
Ave., Livermore, CA 94550, USA.}
\author{Stefan P. Hau-Riege} 
\affiliation{Lawrence Livermore National Laboratory, 7000 East
Ave., Livermore, CA 94550, USA.}
\author{Abraham Sz\"oke}
\affiliation{Lawrence Livermore National Laboratory, 7000 East
Ave., Livermore, CA 94550, USA.}
\author{Congwu Cui}
\affiliation{Advanced Light Source, Lawrence Berkeley National Laboratory, 1 Cyclotron rd. Berkeley, CA 94720, USA}
\author{Malcolm R. Howells}
\affiliation{Advanced Light Source, Lawrence Berkeley National Laboratory, 1 Cyclotron rd. Berkeley, CA 94720, USA}
\author{David A. Shapiro}
\affiliation{Advanced Light Source, Lawrence Berkeley National Laboratory, 1 Cyclotron rd. Berkeley, CA 94720, USA}

\author{John  C. H. Spence}
\affiliation{Department of Physics and Astronomy, Arizona State University, Tempe, AZ 85287-1504, USA}

\author{Joshua W. Shaevitz} 
\affiliation{Department of Physics and Lewis-Sigler Institute, 150 Carl Icahn Laboratory, Princeton, New Jersey 08544, USA.}
\author{Johanna Y. Lee} 
\affiliation{Department of Plant and Microbial Biology, University of California, Berkeley, 648 Stanley Hall 3220, Berkeley, California 94720, USA.}
\author{Janos Hajdu}, 
\affiliation{Stanford Synchrotron Radiation Laboratory, 
Stanford Linear Accelerator Center, 2575 Sand Hill Road, Menlo Park, 
California 94025, USA.}
\affiliation{Laboratory of Molecular Biophysics, 
Department of Cell and Molecular Biology, 
Uppsala University, Husargatan 3, Box 596, SE-75124 Uppsala, Sweden.}
\author{Marvin M. Seibert}
\affiliation{Laboratory of Molecular Biophysics, 
Department of Cell and Molecular Biology, 
Uppsala University, Husargatan 3, Box 596, SE-75124 Uppsala, Sweden.}

\preprint{UCRL-JRNL-234707}

\date{\today}
\maketitle

The ideal microscope for the life and physical sciences should deliver
high-spatial-resolution high-speed snapshots possibly with spectral,
chemical and magnetic sensitivity. X-rays can provide a large
penetration depth, fast time resolution, and strong absorption
contrast across elemental absorption edges. The geometry of Fourier
transform holography is particularly suited to X-ray imaging as the
resolution is determined by the scattering angle, as in
crystallography. The problem with conventional Fourier holography is
an unfavourable trade off between intensity and resolution. We show
here how recent technical developments in lensless flash photography
can be used to reduce this effect.

\begin{figure*}
	\includegraphics[width=0.9\textwidth]{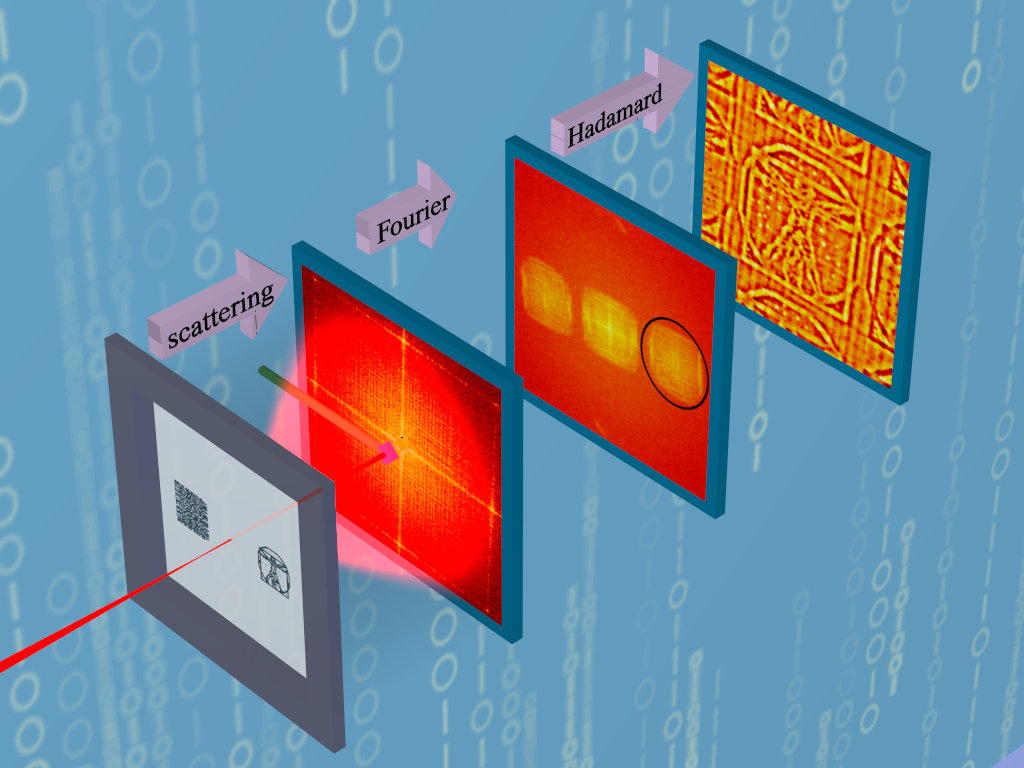}
\caption{
\textbf{Experimental geometry and imaging.}
 A coherent X-ray beam illuminates
both the sample and the Uniformly Redundant Array placed next to
it. An area detector (a Charge Coupled Device in these experiments)
collects the diffracted X-rays. The Fourier transform of the
diffraction pattern yields the autocorrelation map with a holographic
term (in the circle) displaced from the center. The Hadamard transform
decodes the hologram.
\label{fig:1}
}
\end{figure*}

Around the $16^{th}$ century, European
painters used the camera obscura, a dark camera with a pinhole to form
an image on a canvas. Even earlier, pinhole cameras had been used to
image solar eclipses by Chinese, Arab and European
scientists. Scientists and painters knew that a small pinhole was
required for reaching high resolution, but the small pinhole also
dimmed the light and the image\cite{hammond}. 
It was eventually discovered that
lenses could collect a larger amount of light without degrading
resolution. Shutters and stroboscopic flash illumination allowed
recording of the time evolution of the image. The development of the
computer allowed a resurgence of pinhole techniques and random arrays
of pinholes were used, initially in X-ray astronomy
\cite{dicke,abbles}. 
Each bright point of a scene deposits a shadow-image of the pinhole
array on the viewing screen. Depth information about the object is
encoded in the scaling of the shadow image of the object 
points\cite{nugent}. 
Knowledge of the geometry of the pinhole array (the ``coded
aperture'') allowed numerical recovery of the image. Eventually the
pinholes were replaced by binary URAs which were shown to be optimal
for imaging\cite{fenimore:URA}. 
 Their multitudes of sharp features contain equal
amounts of all possible spatial frequencies, thereby allowing high
spatial resolution without sacrifice of image brightness. URA coded
apertures are now commonly used in hard X-ray astronomy\cite{caroli}, medical
imaging\cite{swindell}, plasma research\cite{fenimore:plasma}, 
homeland security\cite{cunningham} and spectroscopy\cite{harwit} to
improve brightness where lenses are not applicable.

The forerunner of our x-ray holography with a
URA reference source is conventional visible-light FTH \cite{stroke}.
 The interference pattern between light scattered by an object and light
from a nearby pinhole is recorded far downstream (Fig. 1). When this
recording (the hologram) is re-illuminated by the pinhole reference
wave, the hologram diffracts the wavefront so as to produce an image
of the object. A second inverted (``twin'') copy of the object is
produced on the opposite side of the optical axis. Under far-field
measurement conditions, a simple inverse-Fourier transform of the FTH
recording produces an image of the specimen convolved with the
reference pinhole source. As in the pinhole image of the camera
obscura, the brightness of the image (or equivalently, the signal to
noise ratio) increases as the reference pinhole increases in size, at
the expense of image resolution.

In general the solution to the
problem of weak signal from a single pinhole (and resulting long
exposure time) is to use multiple reference sources. For example, a
unique mesoscale object has been imaged by x-ray FTH using 5 pinholes
sources\cite{schlotter}. 
No two pairs of pinholes were the same distance apart, so
that each holographic term could be isolated in the autocorrelation
map.  This geometry blocks much of the available light and limits the
number of pixels available to image the specimen. Hitherto efforts to
produce strong signal have been pursued using complicated reference
objects\cite{stroke,szoke,he}.  
 The possible
improvement in signal to noise of Fourier-transform holograms with a
strong reference saturates\cite{collier} at 50\% of that for an ideal 
lens-based amplitude image with loss of phase information.
However there is still the difficulty of deconvolution due to
the missing frequency content of the reference. The flat power
spectrum of the URA is designed to optimize the reference to fill the
detector with light uniformly. In summary, the optimum method of
boosting the holographic signal is to use a URA as the reference
object. The gain in flux compared to a single pinhole is the number of
the opaque elements in the URA (twice as much for phase URAs), in our
case, n=2592 and n=162. The signal to noise ratio (snr) of the
URA-produced image increases initially with the square root of the
number of pixels in the URA\cite{harwit} (see methods), 
in our case by a factor
of 18 and 4.5 for n=2592 and n=162 respectively.

The x-ray FTH experiment is conceptually simple: a coherent beam of
X-rays impinges on a specimen and the coded array, and the diffraction
pattern is recorded far downstream (Fig 1). We report here two
experiments both using an area detector fitted to an existing
experimental end station\cite{beetz,chapman:nphys}, and already used
in several recent coherent X-ray diffractive imaging
experiments\cite{shapiro:pnas,chapman:josaa,chapman:nphys}.  The first
experiment was carried out at beam line 9.0.1 at the Advanced Light
Source (ALS) at the Lawrence Berkeley National Laboratory.  An x-ray
beam defined using a 4 $\mu$m coherence-selecting pinhole was used to
image a test object placed next to a 44 nm resolution array (Figure
2). The flux advantage of the URA method is illustrated by the
calculations shown in Figs. 2 () and (F) and explained in the caption
is made evident from our results.

\begin{figure}
	\includegraphics[width=0.48\textwidth]{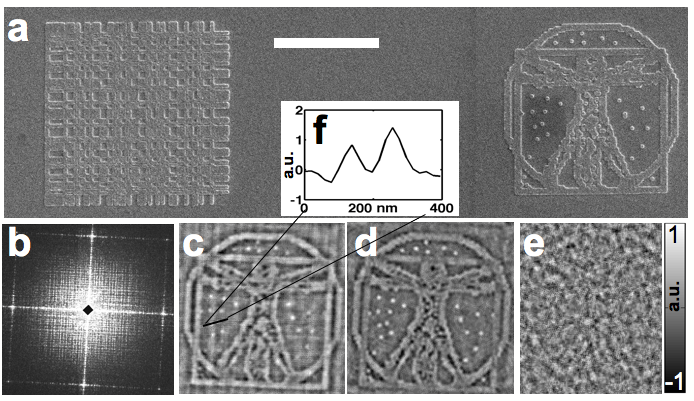}
\caption{
\textbf{High resolution holography.}  
(a) Lithographic test sample next to a 10
nm thick twin-prime 71$\times$73 array of 43.5 nm square gold scattering
elements, imaged by Scanning Electron Microscopy. 
Scale bar is 2 $\mu$m. (b) Diffraction pattern
collected at the Advanced Light Source ($\lambda$=2.3 nm, $10^6$ photons in 5 s exposure, 200
mm from the sample). (c) Real part of the reconstructed hologram
(linear grayscale), the
smallest features of the sample of 43.5 nm are clearly visible. (d)
simulation with $10^6$ photons. (e) simulation with the same number of photons, but a single
reference pinhole. (f)  Cut through two dots separated by 120 nm.
\label{fig:2}
}
\end{figure}

The second series of experiments was carried out at the
FLASH soft X-ray free electron laser ($\lambda$=13.5 nm) in Hamburg. A 15 fs
pulse of $10^{12}$ photons traverses the sample and URA, and is diffracted
just before they both turn into a hot
plasma\cite{chapman:nphys,chapman:nat},
 and become vapourised. 
A bacterium was imaged to demonstrate that the experiment
was feasible with the lower scattering strength of biological material
(Fig. 3).  More information on these experiments is given in the
figure captions and methods section.

In FTH, the final image is
produced by an especially simple one-step calculation.  A Fourier
transformation of the measured intensity pattern delivers the
autocorrelation of the wavefield that exits the object plane.  In
standard FTH with a point reference source such an autocorrelation
would already contain the image.  For FTH with a URA reference source,
it includes the convolution of the object and the URA.  Positioning
the sample and the array with a sufficient spacing between them
ensures that this information-containing term will be separated from
the autocorrelations of the array and the object. To extract the final
image, we convolved the holographic term (with the rest of the
autocorrelation map set to zero) using a mosaic of $3\times3$ 
URAs, 
that is using the same delta-Hadamard transform  used to reconstruct
 the pinhole camera image intensities, but replacing the intensity 
image with the complex valued cross correlation term.
  The reconstructed images made by this method are shown in Figure 2 
using 44-nm URA elements. 
The forward or small angle scattering was discarded during
data collection, yielding edge enhancement in the image. For the
biological image additional refinement was demonstrated.  By using the
Fourier-Hadamard-transform image provided by a 150 nm resolution URA
as the starting point for a phase retrieval algorithm, we refined the
15-fs flash image of a small helical bacterium (Spiroplasma
melliferum) to the full extent of the recorded diffraction pattern at
half the pinhole-size resolution (75 nm) (Fig. 3 and methods
section).

\begin{figure}
	\includegraphics[width=0.48\textwidth]{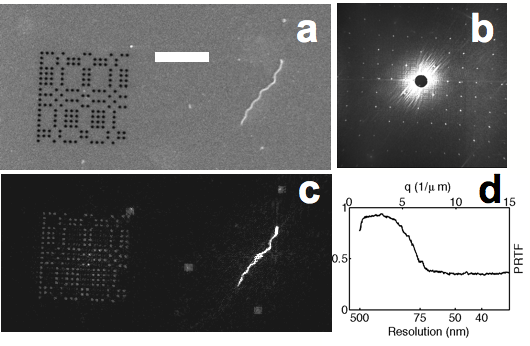}
\caption{
\textbf{Ultrafast imaging and phase extension.} (a) A URA of pinholes is placed
near a Spiroplasma cell (scale bar is 4 $\mu$m). (b) Diffraction pattern
collected at FLASH in a single 15 fs ($\lambda$=13.5 nm) pulse. (c)
Reconstructed image by phase retrieval 
methods\cite{marchesini:rsi,marchesini:prb}. (d) 
the reproducibility of the recovered image as a function of spatial
frequency drops at 75 nm resolution.
\label{fig:3}
}
\end{figure}

The resolution of the holograms corresponds to the
resolution of the fabricated URAs: 44 nm for the lithographic pattern
used at ALS, and 150 nm (refined to 75 nm) for the bacterium used at
FLASH. These values are among the best ever reported for holography of
a one-micron-sized object, and we believe resolution will improve in
the future with the development of nano-arrays. URAs with 25 nm
resolution and 9522 elements have already been produced by
conventional methods.

In conclusion, we have successfully demonstrated holographic X-ray
imaging with URAs and obtained amplified X-ray holographic images at
attractive resolutions. These images were orders of magnitude more
intense than those from conventional Fourier transform holography,
enabling the potential use of novel tabletop sources\cite{wang}.
Since URA diffraction uniformly filled the detector with light, image
reconstruction could be performed by a Fourier-Hadamard transform in a
single step without iterations. The technique shows good prospects for
improvements in the spatial resolution, and the results verify the
predicted high performance values with respect to high brightness and
ultrafast time resolution. Imaging with coherent X-rays will be a key
technique for developing nanoscience and nanotechnology, and or
massively parallel holography will be an enabling tool in this quest.

\acknowledgments
We are grateful to the staff of FLASH
and ALS for help, and to D. A. Fletcher for the Spiroplasma
samples. This work was supported by the following agencies: the
U.S. Department of Energy by Lawrence Livermore National Laboratory
under Contract W-7405-Eng-48 and DE-AC52-07NA27344; the Advanced Light
Source; National Center for Electron Microscopy; Center for X-ray
Optics at Lawrence Berkeley Laboratory under DOE Contract
DE-AC02-05CH11231; the Stanford Linear Accelerator Center under DOE
contract DE-AC02-76-SF00515; the European Union (TUIXS); The Swedish
Research Councils, the DFG-Cluster of Excellence through the
Munich-Centre for Advanced Photonics; the Natural Sciences and
Engineering Research Council of Canada to M. B., and the Sven and
Lilly Lawskis Foundation of Sweden to M.M.S.

\section*{Methods}
\noindent
\textbf{Samples.}
Both ALS and FLASH experiments used samples
microfabricated on silicon nitride membranes supported in a silicon
window frame.
The URA and test pattern for the ALS experiments were
fabricated using Center for X-ray Optics nanowriter, Lawrence Berkeley
National Lab. Polymethyl methacrylate, a positive resist, was
patterned on Si$_3$N$_4$ membrane substrate and subsequently gold
electroplated (10 nm Au). Patterning dose was achieved in a fraction of a second, enabling potential mass production.  

\noindent
Following glutaraldehyde fixation, Spiroplasma cells
were air dried from a solution containing S. melliferum on a Si3N4
window covered by a 10 nm poly-l-lysine. The URA was fabricated next
to a Spiroplasma cell using focused ion beam milling at the National
Center for Electron Microscopy.

\noindent
\textbf{Data.} 
At the ALS, diffaction patterns
were collected using a coherent portion of a soft x-ray beam ($\lambda$=2.3
nm) from an undulator source selected by a 5 $\mu$m pinhole. The hologram
was collected with an in-vacuum back illuminated CCD (1300X1340 20 $\mu$m
pixels) placed at 200.5 mm from the specimen. The direct beam is
blocked by a beamstop placed in front of the detector to prevent
damage to the camera. The beamstop is moved during data collection to
recover a large portion of the diffraction pattern. Total collection
time was 5 seconds. At FLASH the same CCD is placed at 54.9. mm from
the specimen and collects elastically scattered X-rays filtered by a
graded multilayer mirror. A hole in the mirror allows the direct beam
through, removing the need for a beamstop. 

\noindent
\textbf{Reconstruction.}
The autocorrelation map, obtained by Fourier transform of the
diffraction pattern, was multiplied by a binary mask 0 in the region
outside the cross correlation between the object and the URA (the hologram).
). The reconstruction was performed by applying the same processing
used for pinhole camera images, by replacing the recorded intensity
image with the masked complex valued cross correlation term. 
A cyclical convolution with a URA decodes the hologram. 
The URA autocorrelation is a delta function
in periodic or cyclical systems, not when surrounded by empty
space. To mimic the cyclical correlation, a mosaic of 3X3 binary URAs
is convolved with the holographic cross term.
 The reconstruction procedure retrieves the hologram of the
object to a resolution determined by the size the URA elements (at
sub-array spacing resolution). However when the spacing between dots
is larger than the dots, the array no longer scatters optimally the
available light, decreasing the SNR.  
 The signal is proportional to the number of elements in the array, 
and the noise is proportional to the square root of the number of 
elements used to
deconvolve the image. Only 2X2 URAs contribute to the reconstructed
image since the holographic term is at most twice the size as the
array itself (the object is smaller than the array).  The mosaic URAs
are made of +1 and -1 (instead of 1 and 0) terms, yielding an
additional factor of $\sqrt{2}$ the noise. The signal to noise ratio
therefore increases as $snr_n=snr_1 \frac{\sqrt{n}}{2\sqrt{2}}$.  For a phase
URA the signal to noise ratio would increase by a factor of 2.

\noindent
\textbf{Phase retrieval.}
The reference points in the URA and a few dust particles where fixed
in space throughout the optimization process, facilitating phase
retrieval optimization\cite{marchesini:rsi}. The addition of a redundant linear
constraint yields more reliable reconstructions, de facto increasing
the resolution of the retrieved complex valued images. The rest of the
illuminated sample was reconstructed with dynamic support\cite{marchesini:prb}.
 The reproducibility of the image  as a function of spatial frequency\cite{shapiro:pnas} drops at 75 nm resolution.

\end{document}